\documentclass[12pt]{iopart}
\usepackage{graphicx}
\begin{document}

\title[{\footnotesize Systematic Study of High-$p_\mathrm{T}$ Direct
Photon Production with the PHENIX Experiment at RHIC}]{Systematic Study
of High-$p_\mathrm{T}$ Direct Photon Production with the PHENIX
Experiment at RHIC}  

\author{T.~Isobe for the PHENIX\footnote{For the full list of PHENIX
authors and acknowledgements, see Appendix 'Collaborations' of this
volume} Collaboration} 

\address{Graduate School of Science, University of Tokyo, 7-3-1 Hongo,
  Bunkyo-ku, Tokyo 113-0033, Japan}
\ead{isobe@cns.s.u-tokyo.ac.jp}
\begin{abstract}
 When studying the initial state and evolution of the matter created in
 relativistic heavy ion collisions, high-$p_\mathrm{T}$ direct photons
 are a powerful probe.  
 They are created in initial hard processes and in parton fragmentation,
 and possibly in interactions of partons with the hot and dense medium.  
 We present systematic measurements of high-$p_\mathrm{T}$ direct photon
 production in $\sqrt{s_\mathrm{NN}} = 200$~GeV p+p and Au+Au collisions.
 The nuclear modification factor of direct photons
 is shown for $5 < p_\mathrm{T} < 18$~GeV/$c$, and
 at very high transverse momenta it seems to be below unity in the most
 central Au+Au collisions. 
\end{abstract}


\section{Introduction}

Direct photons are a powerful probe to study the initial state of
matter created in relativistic heavy ion collisions since photons do
not interact strongly once produced.  They are emitted in all stages
of the collision: in the initial state where photon production can be
described by NLO pQCD, in the Quark-Gluon Plasma (QGP), dominated by
thermal emission, and in the final hadron-gas
phase~\cite{bib:thermalphoton}. 
In addition, high transverse momentum~($p_\mathrm{T}$) photons are
expected to be produced by the interaction of jet partons with dense
matter. 

The direct photon yields measured in RHIC-Year2 heavy ion collisions by
the PHENIX experiment are in good agreement with a NLO pQCD calculation
scaled by the number of binary nucleon collisions within 
experimental errors and theoretical uncertainties~\cite{bib:photon}.   
While this suggests that the initial hard scattering probability is not 
reduced, the direct photon yield measured in Au+Au collisions is in principle
expected to be suppressed compared to binary scaled NLO pQCD
calculations, because direct photons in the calculation consist of
prompt photons produced directly in hard scattering and jet
fragmentation photons from hard scattered partons which in turn would be
suppressed due to the jet-quenching effect. 
The agreement with NLO pQCD calculations can just be a coincidence
caused by mutually counterbalancing effects of Compton-like scattering
of the jet partons with the medium, often referred to as jet-photon
conversion~\cite{bib:fries} and energy loss of jet partons themselves.  
In order to study this effect for direct photon in the dense matter,
it is important to measure the nuclear modification factor
(R$_\mathrm{AA}$) of direct photon using p+p data as a reference rather
than NLO pQCD calculations as done in earlier publications. 

The PHENIX experiment~\cite{bib:phenix} can measure photons with two
types of highly segmented electromagnetic
calorimeters~(EMCal)~\cite{bib:emcal}.  
One is a lead scintillator sampling calorimeter~(PbSc), and
the other is a lead glass Cherenkov calorimeter~(PbGl).

For the measurement of direct photons presented here, the conventional
subtraction method has been used.
$\pi^0$ and $\eta$ mesons are reconstructed via their two-photon decay
mode. 
The $p_\mathrm{T}$ spectra of direct photons are obtained by subtracting 
the spectra of decay photons from the $p_\mathrm{T}$ spectra of
inclusive photons. 
The decay photons were estimated based on the measured $\pi^0$ and
$\eta$ spectra.  

\section{Result on $\sqrt{s} = 200$~GeV p+p collisions}

\begin{figure}[t]
 \begin{center}
  \begin{minipage}{0.48\linewidth}
   \begin{center}
    \includegraphics[width=\textwidth]{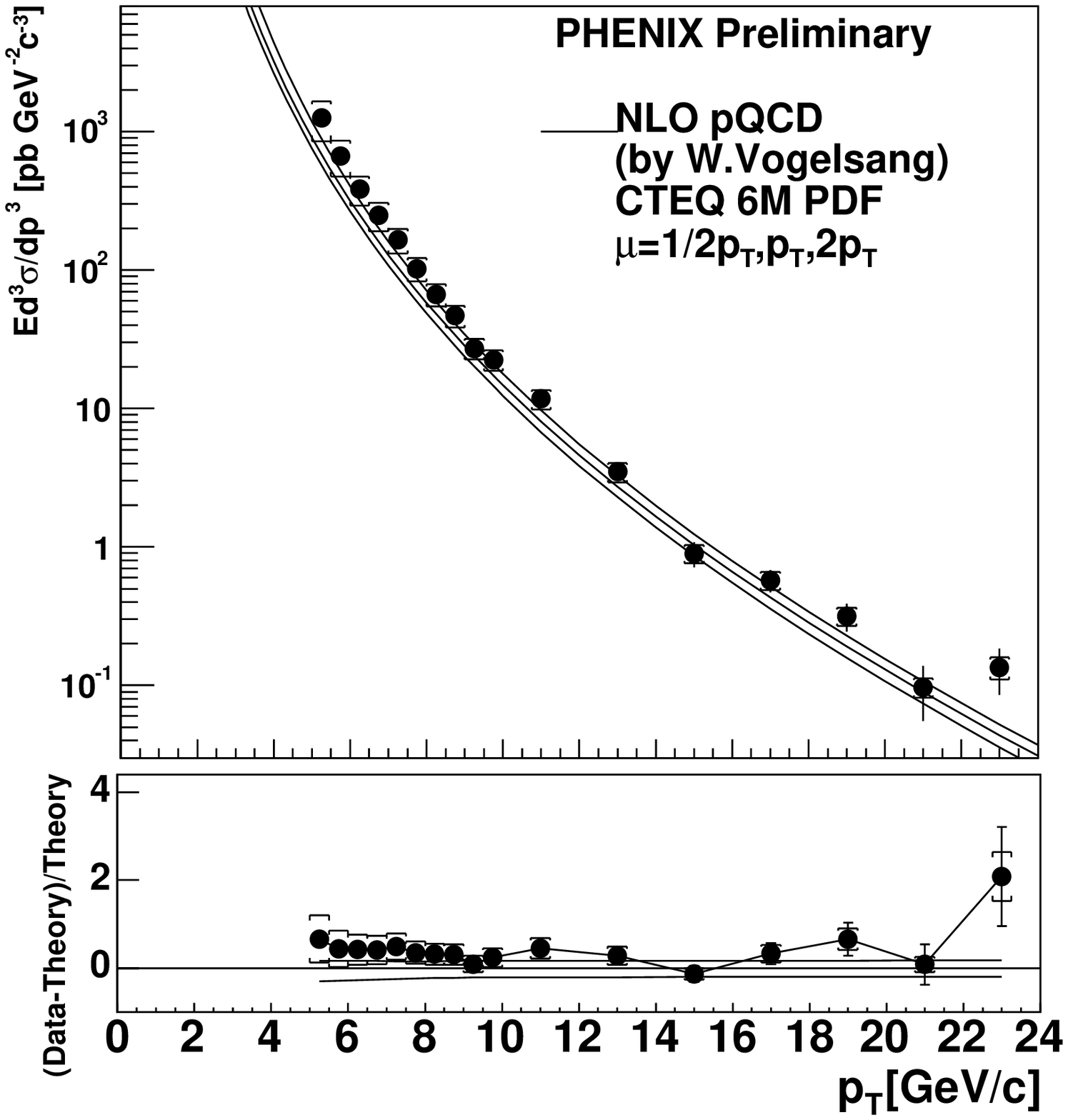}
   \end{center}
  \end{minipage}
  \begin{minipage}{0.48\linewidth}
   \begin{center}
    \includegraphics[width=\textwidth]{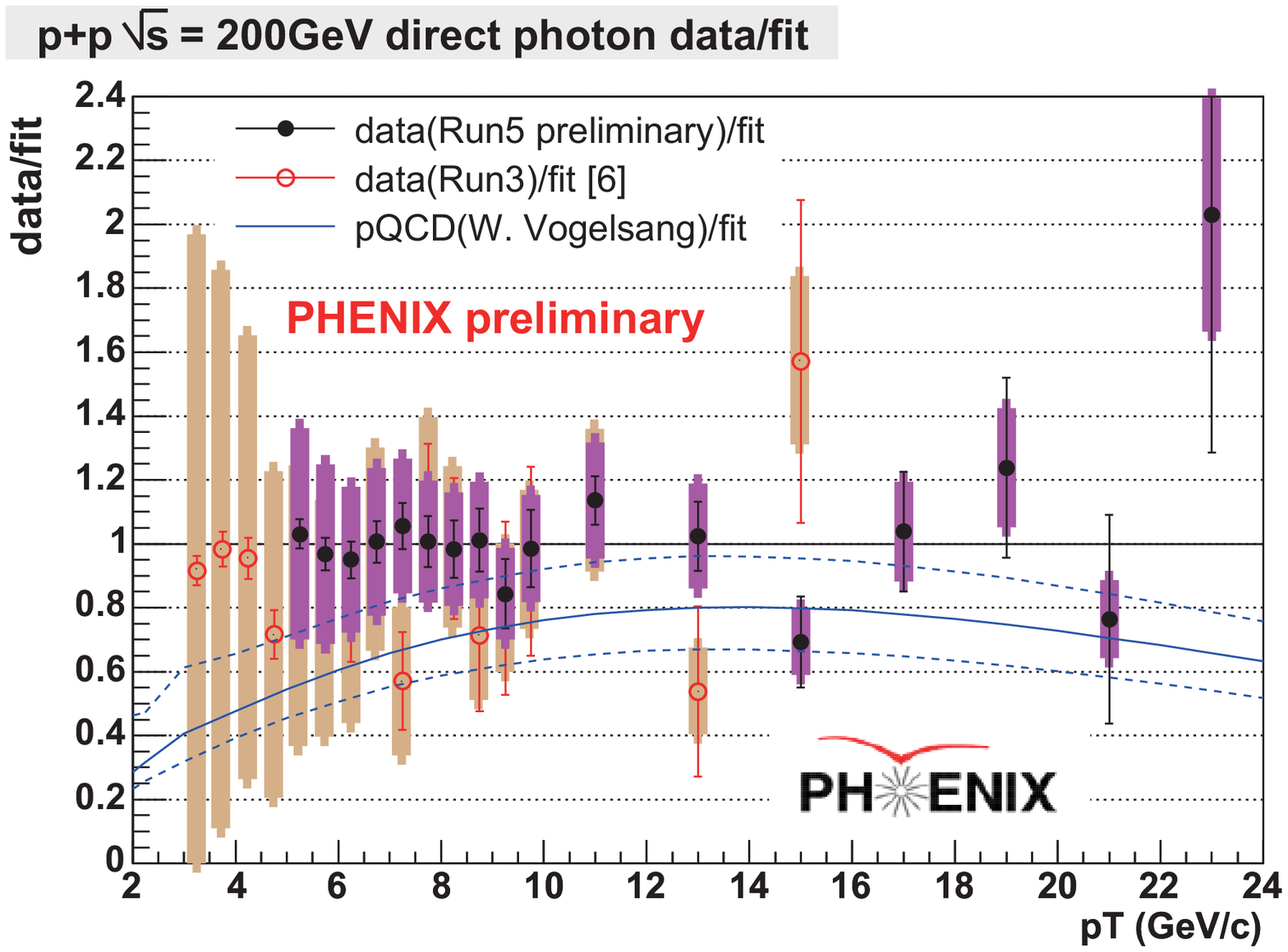}
   \end{center}
  \end{minipage}
 \end{center}
 \caption{Left: Direct photon spectra compared with NLO pQCD
 calculations for three theory scales, $\mu$. Brackets around data
 points show systematic errors. The bottom figure shows the comparison
 to the NLO pQCD calculation for $\mu$ = $p_\mathrm{T}$, with upper and 
 lower curves for $\mu$ = $p_\mathrm{T}$/2 and 2$p_\mathrm{T}$. Right:
 ratio of data and parameterized curve. Bands around data points show
 systematic errors. For comparison, the NLO pQCD calculations are shown
 as well.}     
 \label{fig:ppphoton}
\end{figure}

The direct photon spectra in $\sqrt{s} = 200$~GeV p+p collisions based
upon RHIC-Year3 data have been published in \cite{bib:ppphoton}.  
The new large amount of data recorded by PHENIX in RHIC-Year5 ($\int L$
= 3.8~pb$^{-1}$) makes it possible to extend the $p_\mathrm{T}$ range of
the direct photon cross-section. 
The left panel of Fig.~\ref{fig:ppphoton} shows the preliminary result
for direct photon cross-section in $\sqrt{s} = 200$~GeV p+p collisions.
A NLO pQCD predictions~\cite{bib:pqcd}, using CTEQ 6M parton
distribution functions and the BFG II parton to photon fragmentation
function, with three theory scales ($\mu$) are shown as well.
The data are consistent with the NLO pQCD calculation within the
uncertainties. 

For the measurement of the R$_\mathrm{AA}$ in Au+Au, the direct photon
cross-section in p+p collisions is parameterized.
The right panel of Fig.~\ref{fig:ppphoton} shows how well the
parameterization describes the direct photon cross-section.
The NLO pQCD calculations are also shown.
Although the theoretical calculation can reproduce the experimental data
within the uncertainties, the mean points of the data are systematically
larger than the calculation.  

\section{Result on $\sqrt{s_\mathrm{NN}} = 200$~GeV Au+Au collisions}

\begin{figure}[t]
 \begin{center}
  \begin{minipage}{0.48\linewidth}
   \begin{center}
    \includegraphics[width=\textwidth]{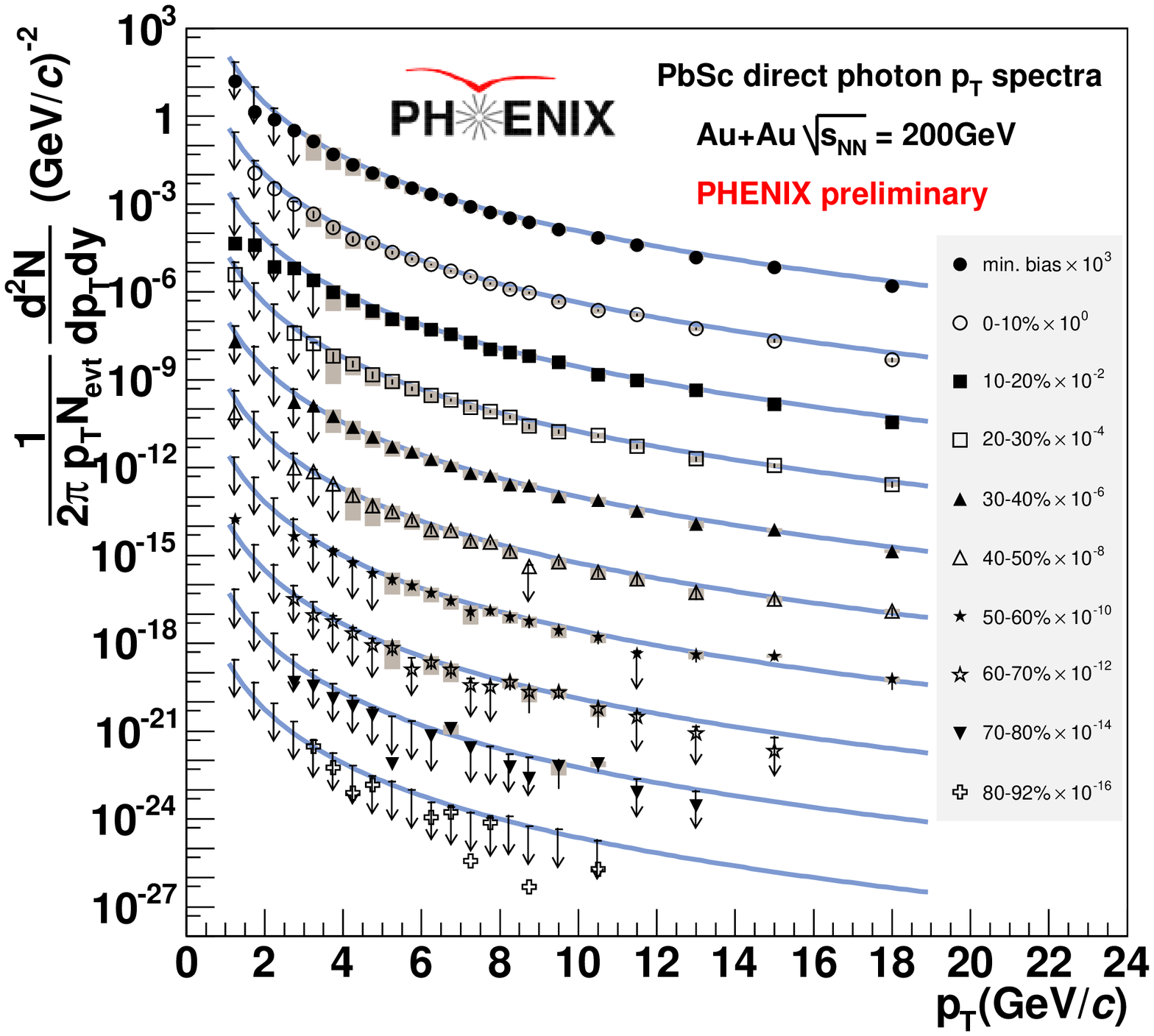}
   \end{center}
  \end{minipage}
  \begin{minipage}{0.48\linewidth}
   \begin{center}
    \includegraphics[width=\textwidth]{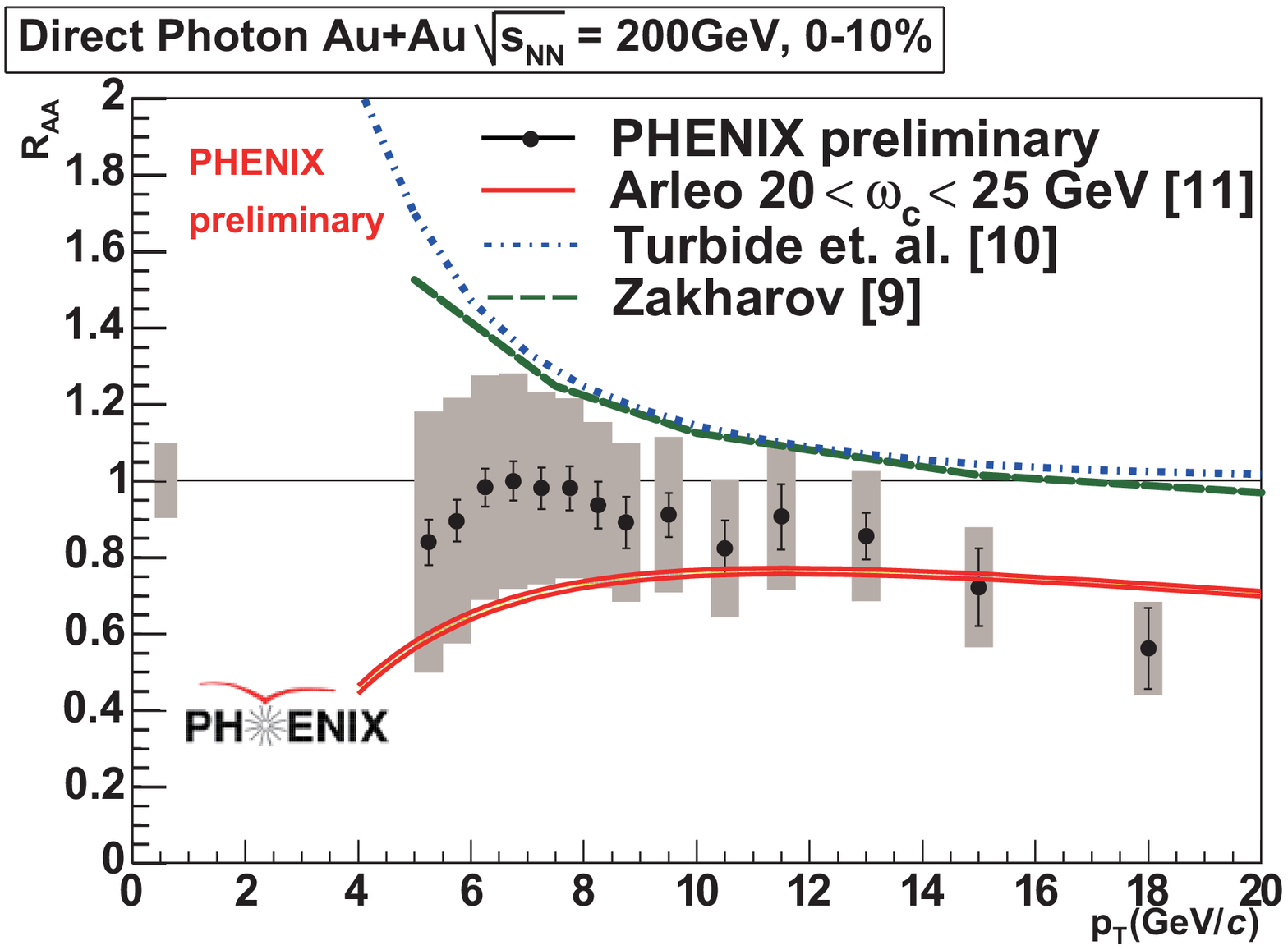}
   \end{center}
  \end{minipage}
 \end{center}
 \caption{Left: direct photon invariant yields as a function of
 $p_\mathrm{T}$ for nine centrality selections and minimum bias Au+Au 
 collisions at $\sqrt{s_\mathrm{NN}} = 200$~GeV.  
 In addition to the statistical and $p_\mathrm{T}$-uncorrelated errors,
 $p_\mathrm{T}$-correlated errors are shown on the data points as bands.
 Arrows indicate measurements consistent with zero yield with the tail
 of the arrow indicating the 90~\% confidence level upper limit. 
 The solid curves are binary scaled p+p direct photon parameterizations. 
 Right: direct photon R$_\mathrm{AA}$ as a function of $p_\mathrm{T}$
 in the most central (0-10~\%) collisions compared with theoretical
 calculations~\cite{bib:zakharov,bib:turbide,bib:arleo}.  
 In addition to the statistical and $p_\mathrm{T}$-uncorrelated errors,
 $p_\mathrm{T}$-correlated errors are shown on the data points
 as boxes, and an overall systematic error of T$_\mathrm{AA}$
 normalization is shown at 1.}
 \label{fig:auauphoton}
\end{figure}

PHENIX recorded high-statistics Au+Au data in RHIC-Year4
($\int L$ = 241~$\mu$b$^{-1}$). 
The new data allow us to measure direct photons and to evaluate their
nuclear modification up to very high-$p_\mathrm{T}$.
Owing to the strong suppression of neutral hadrons in heavy ion
collisions~\cite{bib:pi0}, the signal-to-noise ratio of direct photons
is better for extraction at high-$p_\mathrm{T}$ ($p_\mathrm{T} >
5$~GeV/$c$).  
The direct photon excess ratio ($R =
\gamma_\mathrm{all}/\gamma_\mathrm{bg}$) is about 3 at $p_\mathrm{T} =
10$~GeV/$c$ in the most central (0-10~\%) Au+Au collisions. 
The left panel of Fig.~\ref{fig:auauphoton} shows direct photon spectra
as a function of $p_\mathrm{T}$ for nine centralities and minimum bias
of Au+Au collisions at $\sqrt{s_\mathrm{NN}} = 200$~GeV.    
The spectra shown here are obtained using only the PbSc calorimeter.
The solid curves show the binary scaled fit to the p+p direct photon
data. 

The R$_\mathrm{AA}$ is defined as:

\begin{equation}
 R_\mathrm{AA}(\mathrm{p}_\mathrm{T}) =
  \frac{d^2N^\mathrm{AA}/d\mathrm{p}_\mathrm{T}d\eta}{\left<T_\mathrm{AA}(b)\right>d^2\sigma^\mathrm{pp}/d\mathrm{p}_\mathrm{T}d\eta}, \label{raa} 
\end{equation}

\noindent
where the numerator is invariant direct photon yield in unit rapidity in
Au+Au collisions and the denominator is the expected yield from p+p
collisions scaled by the nuclear overlapping
function~($\left<T_\mathrm{AA}(b)\right> =
\left<N_\mathrm{coll}(b)\right>/\sigma_\mathrm{pp}$) in Au+Au.    
If the reaction is just the superposition of hard scatterings,
unmodified by nuclear effects, the R$_\mathrm{AA}$ is
unity. 

The right panel of Fig.~\ref{fig:auauphoton} shows R$_\mathrm{AA}$
of direct photons using the p+p reference as a function of
$p_\mathrm{T}$ in the most central (0-10~\%) Au+Au collisions at
$\sqrt{s_\mathrm{NN}} = 200$~GeV, in comparison to several theoretical
expectations~\cite{bib:zakharov,bib:turbide,bib:arleo}.    
R$_\mathrm{AA}$ of direct photons seems to be below unity at very
high-$p_\mathrm{T}$ ($p_\mathrm{T} > 14$~GeV/$c$).
None of the theoretical calculations reproduces R$_\mathrm{AA}$ in the
full $p_\mathrm{T}$ range. 
Jet-photon conversion is not taken into account in Arleo's
calculation~\cite{bib:arleo}, and so-called isospin effect is not taken
into account in Turbide {\it et al.}'s calculation~\cite{bib:turbide}.  
In order to measure R$_\mathrm{AA}$ of direct photons more precisely at
very high-$p_\mathrm{T}$, it is necessary to measure high-$p_\mathrm{T}$
direct photon with the PbGl calorimeter, which can provide same
measurement with different systematics that would give more detailed
insight of the high-$p_\mathrm{T}$ direct photons.

\section{Summary}
The measurements of high-$p_\mathrm{T}$ direct photons in p+p and Au+Au
collisions at $\sqrt{s_\mathrm{NN}} = 200$~GeV are presented.
The direct photon signal is extracted as a function of the Au+Au
collision centrality and R$_\mathrm{AA}$ is calculated with
p+p direct photon data.
The R$_\mathrm{AA}$ of very high-$p_\mathrm{T}$ direct photon seems to
be below unity in the most central collisions.

\end{document}